\documentclass[apj]{emulateapj}

\begin{document}

\title{The Redshift Distribution of Giant Arcs in the Sloan Giant Arcs Survey}

\author{
Matthew B. Bayliss\altaffilmark{1,2}, 
Michael D. Gladders\altaffilmark{1,2},
Masamune Oguri\altaffilmark{3},
Joseph F. Hennawi\altaffilmark{4}, 
Keren Sharon\altaffilmark{2},
Benjamin P. Koester\altaffilmark{1,2}, 
H{\aa}kon Dahle\altaffilmark{5} 
}

\email{mbayliss@oddjob.uchicago.edu}

\altaffiltext{1}{Department of Astronomy \& Astrophysics, University of Chicago, 5640 South Ellis Avenue, Chicago, IL 60637}
\altaffiltext{2}{Kavli Institute for Cosmological Physics, University of Chicago, 5640 South Ellis Avenue, Chicago, IL 60637}
\altaffiltext{3}{Division of Theoretical Astronomy, National Astronomical Observatory of Japan, 2-21-1 Osawa, Mitako, Tokyo 181-8588, Japan}
\altaffiltext{4}{Max-Planck-Institut f\''{u}r Astronomie K\''{o}nigstuhl 17, D-69117, Heidelberg, Germany}
\altaffiltext{5}{Institute of Theoretical Astrophysics, University of Oslo, P.O. Box 1029, Blindern, N-0315 Oslo, Norway}

\begin{abstract}

We measure the redshift distribution of a sample of $28$ giant arcs discovered 
as a part of the Sloan Giant Arcs Survey (SGAS). Gemini/GMOS-North 
spectroscopy provides precise redshifts for $24$ arcs, and ``redshift 
desert'' constraints for the remaining $4$. This is a direct measurement 
of the redshift distribution of a uniformly selected sample of bright giant 
arcs, which is an observable that can be used to inform efforts to predict 
giant arc statistics. Our primary giant arc sample has a median redshift 
$z=1.821$ and nearly two thirds of the arcs -- $64\%$ -- are sources at 
$z \gtrsim 1.4$, indicating that the population of background sources that 
are strongly lensed into bright giant arcs resides primarily at high redshift. 
We also analyze the distribution of redshifts for $19$ secondary strongly 
lensed background sources that are not visually apparent in SDSS imaging, 
but were identified in deeper follow-up imaging of the lensing cluster fields. 
Our redshift sample for the secondary sources is not spectroscopically 
complete, but combining it with our primary giant arc sample suggests that 
a large fraction of $all$ background galaxies which are strongly lensed by 
foreground clusters reside at $z\gtrsim1.4$. Kolmogorov-Smirnov (KS) tests 
indicate that our well-selected, spectroscopically complete primary giant arc 
redshift sample can be reproduced with a model distribution that is constructed 
from a combination of results from studies of strong lensing clusters in numerical
simulations, and observational constraints on the galaxy luminosity function.

\end{abstract}


\section{Introduction}

Comparisons of the observed giant arc counts against 
theoretical predictions provide a test for cosmological models of 
structure formation. Giant arc statistics depend on the growth of 
structure through the abundance and internal properties of the most massive galaxy 
clusters that dominate the giant arc cross-section. \citet{bartelmann1998} 
suggested that there was an apparent order-of-magnitude discrepancy between 
the observed counts of giant arcs on the sky, and what is predicted by 
a $\Lambda$CDM cosmology. Subsequent comparisons of homogeneous samples of giant arcs 
against theoretical predictions have provided additional evidence for an apparent 
``giant arc problem'' \citep{gladders2003,zaritsky2003}. \citet{gladders2003} present 
a sample of five giant arcs that are identified in the Red-Sequence Cluster Survey (RCS), 
and \citet{zaritsky2003} identified three strongly lensed systems in the Las Campanas 
Distant Cluster Survey (LCDCS), but both of these samples are very small and are 
therefore subject to prohibitively large uncertainties due to small number statistics.

Further studies have explored a variety of potential factors that could 
help to explain the underabundance of giant arcs predicted by theoretical models, 
such as accounting for galaxies by painting them onto simulated dark matter 
halos \citep{flores2000,meneghetti2000,puch2009}, steepening of the gravitational 
potential in cluster cores due to baryonic dissipation effects \citep[e.g.,][]{puch2005,rozo2008}, 
contributions from secondary structures along the line of sight \citep{puch2009}, 
accounting for short time-scale increases in the lensing cross-sections of halos 
due to mergers \citep{torri2004}, and varying the 
redshift distribution of the population of background sources available to be 
strongly lensed into giant arcs \citep{hamana1997,oguri2003,wambs2004}. 
\citet{hennawi2007} found this last factor -- the background source 
distribution -- to be the one of the leading sources of uncertainty in their 
efforts to produce precise estimates for giant arc statistics from ray-tracing 
of dark matter simulated halos for a given cosmology. The primary motivation
for studying giant arc statistics in a cosmological context is to provide
a comprehensive test for models of large scale structure, which require 
$a$~$priori$ constraints other model inputs, such as the properties of the 
background source population. Fortunately, the redshift distribution of giant 
arcs is directly observable given a sufficiently large and well-selected
sample of giant arcs with redshift measurements. Samples of giant arc redshifts
exist in the literature, including the catalog compiled by \citet{sand2005} and the
redshifts used by \citet{richard2009}, but they cannot be used to study the true 
redshift distribution of arcs because of their non-uniform selection and 
severe spectroscopic incompleteness.

In this letter we measure the redshift distribution of a spectroscopically complete 
sample of $28$ giant arcs that were discovered in the Sloan Giant Arc Survey (SGAS; M.~D. 
Gladders et al., in preparation) and targeted for multi-object spectroscopy with 
Gemini/GMOS-North. We also consider a secondary sample of redshifts that were 
measured for an additional $19$ strongly lensed sources. For comparison we test the 
redshift distributions of our primary and secondary samples against simple models for 
the redshift distribution of giant arcs. 

All magnitudes given in this letter are AB, calibrated relative to the SDSS 
\citep{york2000}.

\section{Data and Samples}

\subsection{Giant Arc Redshift Sample}

The lensed background sources discussed in this letter are located near 
the cores of foreground massive galaxy clusters with a median redshift, 
$z_{l}=0.435$, and a median dynamical mass, $M_{Vir} = 7.84\times10^{14} 
M_{\sun} h_{0.7}^{-1}$ \citep{bayliss2010a}. 

We take published spectroscopic 
redshifts from \citet{bayliss2010a}, and use optical colors combined with the 
absence of specific spectral features to identify upper and lower redshift bounds 
for some arcs that lack precise spectroscopic redshifts. Given the spectral 
coverage of the Gemini/GMOS spectroscopy presented in \citet{bayliss2010a} 
we expect to observe one or more prominent emission lines 
(e.g. O[II]~$\lambda3727$\AA, H-$\beta~\lambda4861$\AA, 
O[III]~$\lambda4959,5007$\AA~ and H-$\alpha~\lambda6563$\AA) for galaxies at 
$z\lesssim1.5$ that are actively star-forming, with some slight variation in this 
redshift limit that depends on the exact spectral coverage for each source. 
For strongly lensed sources at $1.5\lesssim z \lesssim 3.3$ we must rely on rest-frame 
UV features to measure redshifts, but the strength of these features can vary 
significantly depending on the physical properties of the individual galaxies. 
At redshifts of $z \gtrsim 3.3$ we expect to see a broadband flux decrement that 
would be indicative of the Lyman-$\alpha$ Forest and Lyman Limit being redshifted 
well into the $g-$band, as well as strong absorption or emission from 
Lyman-$\alpha~\lambda1216$\AA~ redshifted into our data for sources with 
spectral coverage extending sufficiently blueward.

Using the criteria outlined above we identifty six arcs in the 
\citet{bayliss2010a} Gemini/GMOS data for which we can confidently place
lower and upper limits on the redshifts. These six arcs all have blue colors 
in the available photometry, indicating that they are actively star-forming, and 
their optical spectra consist of blue continuum emission with no strong absorption 
or emission features. The precise redshift constraints vary slightly from arc to 
arc, depending on the exact limits of the wavelength coverage, and are given in 
Table~\ref{tabzdesert}.

\begin{deluxetable*}{lllllll}
\tablecaption{Redshift Constraints from Gemini Spectroscopy\label{tabzdesert}}
\tablewidth{0pt}
\tabletypesize{\tiny}
\tablehead{
\colhead{Cluster Core} &
\colhead{Label\tablenotemark{a}} &
\colhead{$z_{arc}$ range} &
\colhead{$l/w$\tablenotemark{b}} &
\colhead{R$_{arc}$\tablenotemark{c}} &
\colhead{AB Mag\tablenotemark{d}} &
\colhead{Source Type} }
\startdata
SDSS J1028+1324 & C & $1.58\leq z \leq3.3$ & $14$ & $17\arcsec$ & $g=22.81$ & Primary \\
SDSS J1115+5319\tablenotemark{e} & A & $1.45\leq z \leq3.2$ & $10$ & $31\arcsec$ & $g=23.03$ & Secondary \\
SDSS J1152+0930 & D & $1.56\leq z \leq3.3$ & $18$ & $14\arcsec$ & $g=22.54$ & Primary \\
GHO 132029+315500\tablenotemark{e} & A & $1.5\leq z \leq3.3$ & $20$ & $21\arcsec$ & $g=22.59$ & Secondary  \\
SDSS J1446+3033 & F & $1.45\leq z \leq3.3$ & $7$ & $15\arcsec$ & $g=22.42$ & Primary \\
SDSS J1456+5702 & B & $1.55\leq z \leq3.3$ & $7$ & $17\arcsec$ & $g=20.83$ & Primary \\
\enddata
\tablenotetext{a}{~Labels identify arcs in figures and tables in \citet{bayliss2010a}}
\tablenotetext{b}{~Length-to-width ratios are all estimated from ground-based imaging with variable seeing, and generally represent lower limits on the true l/w for each arc. In the case of multiple arcs/images of a single source, the component with the largest length-to-width ratio is reported.}
\tablenotetext{c}{~R$_{arc}$ is measured by calculating the mean distance from a giant arc to the BCG of the lensing cluster.}
\tablenotetext{d}{~Magnitudes are integrated total aperture magnitudes for the brightest contiguous piece of each arc.}
\tablenotetext{e}{~These arcs were not identified in visual inspection of SDSS survey imaging, but rather in deeper imaging of red sequence selected galaxy clusters.}
\end{deluxetable*}

\begin{figure}
\includegraphics[scale=0.58]{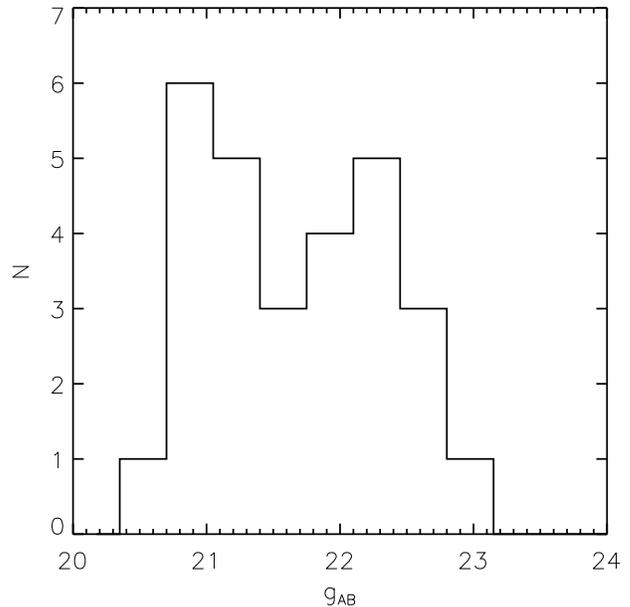}
\caption{\scriptsize{
Histogram of $g-$band magnitudes for the $28$ primary giant 
arcs discussed in this letter.}}
\label{arcmags}
\end{figure}

\subsection{Primary Giant Arc Sample}

Arcs in our primary giant arc sample meet two criteria: 1) they 
were visually identified in the original SDSS imaging data $and$ 2) they were 
were observed spectroscopically with Gemini/GMOS-North. Our primary giant arcs span 
a range in integrated $g-$band magnitude of $20.69 \leq g \leq 22.81$ 
(Figure~\ref{arcmags}). This sample includes all sources identified as primary giant 
arcs in \citet{bayliss2010a}, plus four additional arcs with redshift desert constraints 
as described in Section 2.1 of this letter (``Primary'' in Table~\ref{tabzdesert}). 
We are spectroscopically complete for the entire primary sample.

\subsection{Secondary Background Source Sample}

Our sample of secondary background sources includes objects that appear to 
be strongly lensed but lack the brightness and/or morphology to be identified in 
the visual selection processes that produced our primary giant arc sample. Most of 
these sources were identified in the Gemini/GMOS pre-imaging data described in 
\citep{bayliss2010a} and are designated as ``secondary strongly lensed sources'' 
in that paper. The secondary sample selection is not as uniform as the primary 
sample, and is spectroscopically incomplete, as there were many of these candidate 
strongly lensed sources that were targeted in the Gemini observations of 
\citet{bayliss2010a} but for which the data do not provide a redshift. 

\begin{figure}
\centering
\includegraphics[scale=0.57]{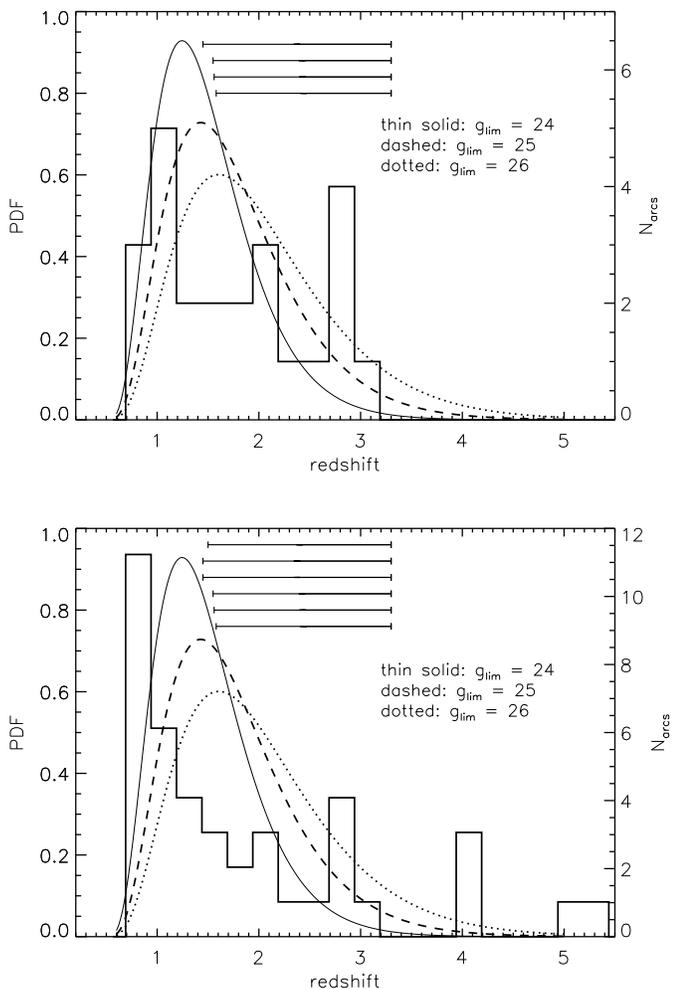}
\caption{\scriptsize{
$Top:$ The distribution of redshifts for the primary giant 
arc sample is plotted as the solid histogram with binsize 
$\Delta z = 0.25$. Arcs with redshift desert constraints are 
plotted as horizontal error bars indicating the upper and lower 
redshift limits for each. Over-plotted are three models for 
the expected redshift distribution assuming different intrinsic 
limiting magnitudes, as described in Section 3.1, all 
of which have a density profile slope, $\alpha=-1.5$. 
$Bottom:$ The same models as in the top panel, plotted with the 
redshift distribution for our giant arc plus secondary source 
combined sample. This combined sample redshift distribution is 
not compatible with any of the model redshift distributions.
}}
\label{arcz}
\end{figure}

\section{The Redshift Distribution of Giant Arcs}

\subsection{Models}

Efforts to produce theoretical predictions for the abundance of giant arcs have modeled 
the population of background source galaxies in a variety of ways. It is known that the 
redshift distribution of background sources can have a dramatic impact on the giant arc 
statistics produced for a given cosmology \citep{hamana1997,oguri2003,wambs2004}. Even 
more problematic is the fact that the uncertainty in the redshift distribution of 
background sources available to be lensed is effectively degenerate with the integrated 
lensing cross-section of the foreground halo population, and therefore with key cosmological 
parameters that strongly impact the properties of halos. This degeneracy arises because 
the Einstein radius, $\theta_{E}$, of a lens at a fixed redshift increases as the distance 
to the source plane increases through its dependence on the angular diameter distance to 
the source.

The simplest background source model is one that places all background sources 
at a common redshift \citep[e.g.,][]{bartelmann1998,meneghetti2010}, for example 
$z_{s}=1$ or $z_{s}=2$, with some number density on the sky. Efforts by some 
groups \citep{wambs2004,dalal2004,hennawi2007,puch2009,oguri2009} used a more 
advanced approach with background galaxies placed at several discrete source 
planes, incorporating empirical measurement of the density of galaxies 
in different redshift bins in order to capture the evolution of galaxy counts. 
The galaxy counts are typically taken from deep pencil-beam surveys, and 
\citet{hennawi2007} point out that cosmic variance in such survey fields can be as 
large as $50\%$, with measurements of the galaxy density on the sky in different 
surveys differing by more than a factor of $2$ in the same redshift bin. This 
discretized source plane model is an improvement relative to the single source plane 
approach, but still falls short of being physically realistic. \citet{horesh2005} 
simulated giant arc statistics using the same clusters in \citet{bartelmann1998} 
and use a background source distribution that is constructed entirely from 
photometric redshifts in the Hubble Deep Field (HDF). This model is encouraging 
in its use of a smooth redshift distribution, but the HDF is only $5$ arcmin$^{2}$ 
and suffers from dramatic cosmic variance uncertainties \citep{hennawi2007}.

We construct a new model that attempts to predict the redshift distribution of 
giant arcs as function of the limiting $intrinsic$ brightness of background 
sources that are distorted into detectable giant arcs. This is equivalent 
to setting a limiting magnitude in giant arc apparent brightness of  and 
assuming a uniform magnification boost for all sources. In reality magnification 
factors vary for different sources depending on the details of each lens-source 
interaction. Assuming typical total magnifications of $\sim10-25$, and noting an 
approximate integrated limiting magnitude of $g\lesssim22.5$ for our primary giant 
arc sample, we produce model giant arc redshift distributions for two cases: 1) an 
intrinsic limiting source magnitude, $g_{lim} < 25$ and 2) $g_{lim} < 26$. We also 
produce a model redshift distribution for an instrinsic limiting source magnitude of 
$g_{lim} < 24$, which has a peak in the probability distribution closer to $z_{s}=1$.

Our model is calculated as follows: for the arc cross-section, $\sigma_{\rm arc}$, 
we adopt the scaling relation derived by \citet{fedeli2010}, which is based on large 
volume N-body simulations. The \citet{fedeli2010} cross-section is defined from 
simulated arcs with length-to-width ratio, $l/w \gtrsim 7.5$, all of which are produced 
from sources at a redshift, $z_{s}=2$. We derive the arc cross-sections for different 
source redshifts assuming a universal matter density profile slope, $\alpha$, 
in the region around the Einstein radii for cluster lenses. The Einstein radius 
is always located in the very center of the lensing cluster potential where we 
expect a profile slope of $\sim -1.5$ \citep{moore1999}. In order to account for 
some uncertainty in the exact slope of the density profile in the core we 
produce models evaulated for a small range of values for the slope, 
$\alpha = (-1.7, -1.5, -1.3)$.

The redshift distribution of source galaxies for a given magnitude limit, $dn/dz_{s}$, is 
estimated using the photometric redshift catalog in the $2$ deg$^{2}$ COSMOS 
field \citep{ilbert2009}. We then compute the expected redshift distribution 
of giant arcs by the equation:

$$\frac{dp_{\rm arc}}{dz_{s}} = \frac{\sigma_{\rm arc} \frac{dn}{dz_{s}}}{\int dz_{s} \sigma_{\rm arc} \frac{dn}{dz_{s}}}$$ 

The length-to-width limit used to define the arc cross section scaling relation in 
\citet{fedeli2010} is consistent with the length-to-width lower limits on our 
sample of giant arcs as measured from ground-based imaging data \citep{bayliss2010a}. 

\begin{deluxetable*}{ccccc}
\tablecaption{Kolmogorov-Smirnov Probabilities for Models Redshift Distributions\label{kstests}}
\tablewidth{0pt}
\tabletypesize{\tiny}
\tablehead{
\colhead{Redshift Desert\tablenotemark{a}} &
\colhead{$\alpha$\tablenotemark{b}} &
\colhead{$g < 24$} &
\colhead{$g < 25$} &
\colhead{$g < 26$} }
\startdata
\cutinhead{Primary Giant Arc Sample}
monte carlo  &  $-1.7$  &  0.001  &  0.046  &  0.500 \\
monte carlo  &  $-1.5$  &  0.003  &  0.165  &  0.251 \\
monte carlo  &  $-1.3$  &  0.039  &  0.096  &  0.021 \\
minimum  &  $-1.7$  &  0.020  &  0.457  &  0.581 \\
minimum  &  $-1.5$  &  0.061  &  0.738  &  0.128 \\
minimum  &  $-1.3$  &  0.404  &  0.072  &  0.002 \\
maximum  &  $-1.7$  &  8.34e-5  &  0.010  &  0.201 \\
maximum  &  $-1.5$  &  3.29e-4  &  0.050  &  0.251 \\
maximum  &  $-1.3$  &  0.008  &  0.096  &  0.021 \\
\cutinhead{Primary + Secondary Arc Samples}
monte carlo  &  $-1.7$  &  3.57e-4  &  0.039  &  0.014  \\
monte carlo  &  $-1.5$  &  0.001  &  0.013  &  0.0021  \\
monte carlo  &  $-1.3$  &  0.004  &  4.53e-4  &   1.58e-5 \\
minimum  &  $-1.7$  &  0.010  &  0.094  &   0.014 \\
minimum  &  $-1.5$  &  0.033  &  0.013  &  0.001  \\
minimum  &  $-1.3$  &  0.004  &  1.48e-4  &  9.22e-8  \\
maximum  &  $-1.7$  &  6.26e-6  &  8.44e-4  &  0.014  \\
maximum  &  $-1.5$  &  4.05e-5  &  0.002  &  0.002  \\
maximum  &  $-1.3$  &  2.95e-4  &  4.53e-4  &  1.71e-5  \\
\enddata
\tablenotetext{a}{~We handle arcs redshift desert constraints in three ways: ``Monte Carlo'' -- redshift desert arcs are given precise redshifts drawn randomly from the sample of giant arc redshifts that lay within their respective upper and lower redshift bounds; ``minimum'' -- all redshift desert sources are assigned redshifts equal to their lower redshift bound; ``maximum'' -- all redshift desert sources are assigned redshifts equal to their upper redshift bound.}
\tablenotetext{b}{~Slope of the density profile used.}
\end{deluxetable*}

\subsection{Properties of the SGAS Giant Arc Redshift Distribution}

The redshift distribution for our primary giant arc sample is shown in 
Figure~\ref{arcz}, along with three curves which demonstrate the kinds of 
giant arc redshift distributions predicted by the model described above. 
Figure~\ref{arcz} also includes a plot of our 
primary giant arc redshift sample combined with our spectroscopically incomplete 
sample of secondary source redshifts. We use the KS test to quantify the agreement 
between our measured redshift distributions and our proposed models. Conducting KS 
tests on our data requires that we correctly account for the giant arcs in our sample 
with constraints that place them in the redshift desert. To do this we use the information 
that we have about the distribution of giant arcs within the redshift desert: the 
precise redshifts that we measure for $13-14$ giant arcs that are located within 
the redshift desert (the number depends on the exact redshift desert constraints). 

We produce many Monte Carlo realizations of our redshift distribution by assigning 
a precise redshift for each redshift desert arc that is drawn randomly from our sample 
of giant arcs that $do$ have precise redshifts within the corresponding upper and 
lower redshift bounds. This method assumes that our giant arcs with precise redshifts 
between $1.45 \lesssim z \lesssim 3.3$ are distributed within that range in the same 
way as our giant arcs that have only redshift desert constraints. We have no strong 
evidence to contradict this assumption and we consider it to be the most robust method 
for including the giant arcs with upper and lower redshift bounds in the KS tests. 
Realizations that draw precise redshifts at random from within the upper and lower 
bounds for a given redshift desert arc produce distributions that are indistinguishable 
from the realizations that draw redshifts from our sample within those bounds, indicating 
that giant arcs with precise redshifts that fall within the redshift desert, 
$1.5 \lesssim z \lesssim 3.3$, are evenly distributed within that range.

We also explore alternative ways of handling those arcs with redshift desert constraints 
in order to explore how the two extreme possible deviations from the Monte Carlo realization 
method described above affect the results of the KS tests. The first extreme corresponds 
to the case where all redshift desert arcs have true redshifts at or near their lower 
redshift bound, and the opposite extreme where all redshift desert arcs have true redshifts 
that are at or near their upper redshift bound. This gives us three different giant arc 
redshift samples that span the range of possible true redshift distributions for our complete 
primary giant arc sample.

The realizations for our primary giant arc sample are tested against our models 
to produce average KS probabilities of the likelihood that our redshift data are 
drawn from the model distributions. We perform another series of KS tests on the 
combined primary giant arc plus secondary source samples, where the two secondary 
arcs with redshift desert constraints are handled in exactly the same three ways 
as the redshift desert primary giant arcs. Results from all KS tests are shown in 
Table~\ref{kstests}, and from these results we infer that the redshift distribution 
of our giant arc sample is reasonably well described by the our simple arc redshift 
distribution model, with values for the slope of the density profile, $\alpha$ 
that agree with the CDM paradigm and with an intrinsic limiting magnitude in the 
$g$-band that corresponds to a reasonable average magnification for each arc, 
$\sim10-25\times$. 

The best agreement between our giant arc redshift sample and the model redshift 
distributions occurs when the arcs with redshift desert constraints are assumed 
to have true redshifts at or near their lower redshift bound; in this case our 
data have a $\sim74\%$ chance of being drawn from the same probabilty distribution 
as the model corresponding to a limiting intrinsic source mangitude of $g_{lim}<25$ 
and a density profile with slope $\alpha = -1.5$. It is reasonable to expect our 
arcs in the redshift desert to be more likely to have true redshifts closer to 
the lower end of the allowable range because of the spectral features that fall 
into our observed wavelength range. For example, at $1.6 \lesssim z \lesssim 2.3$ 
we must rely on lines such as MgII~$\lambda2796,2803$\AA, and 
FeII~$\lambda2344,2372,2384,2586,2600$\AA, which tend to be relatively weak compared 
to bluer features, such as CIV~$\lambda1448,1551$\AA, SiII~$\lambda1260,1527$\AA, and 
SiIV~$\lambda1394,1403$\AA. The combined primary $+$ secondary strongly lensed source 
sample is never in good agreement with our model redshift distributions, which 
underscores the importance of using well-selected and spectroscopically complete 
redshift samples to study giant arc redshift distributions.

It is possible that our giant arc redshift distribution could be biased high relative 
to the true redshift distribution of SGAS giant arcs due to a systematic selection 
effect. Targets were selected for Gemini spectroscopy with a preference toward 
systems with larger apparent arc radii, R$_{arc}$, which is measured as the average 
angular separation between an arc and the center of the cluster lens -- typically 
coincident with the brightest cluster galaxy. A selection based on large R$_{arc}$ 
could bias our giant arcs toward having higher redshifts than if we had selected 
targets completely at random. However, our spectroscopic target selection was 
not based purely on R$_{arc}$, and as discussed in \citet{bayliss2010a} we expect 
that it does not have a strong impact on our results.

\section{Summary and Conclusions}

From the comparison of our redshift data against models for possible background source 
distributions we conclude that our data are consistent with a model that combines 
strong lensing cross-sections derived from simulations, galaxy counts from COSMOS, 
and the approximate brightness limit of our giant arc sample. Our highest KS test 
probability of $\sim74\%$ occurs when we assume that the redshift desert sources have 
redshifts equal to their lower redshift bound and use a model with a limiting source 
magnitude of $g_{lim} < 25$ and a profile slope $\alpha = -1.5$. Models with 
$g_{lim} < 26$, and $\alpha = -1.7$ also produce KS test probabilities of $\gtrsim50\%$ 
when we assign the redshift desert souces as having redshifts equal to their lower 
bound and also when we use Monte Carlo random sampling to assign redshifts to redshift 
desert sources. Our primary giant arcs have a median redshift of $z=1.821$, and 
approximately $64\%$ (18/28) of the primary giant arcs are located at $z\gtrsim1.45$, 
indicating that the background galaxies which are strongly lensed into bright giant 
arcs tend to reside at high redshift. This result is encouraging for future prospects 
to discover very large samples of bright, strongly lensed galaxies at high redshift 
that can be targeted for detailed individual study.

The analysis presented here also represents an important step forward for studying 
the number and distribution of giant arcs produced by strong lensing clusters. Combinging 
our median giant redshift of $z_{s}=1.821$ with the results from \citet{wambs2004} 
indicates that the ``giant arc problem'' identified by \citet{bartelmann1998} may be 
partially due to the authors' decision to make the simple assumption that all giant arcs are 
background sources at $z_{s}=1$. However, we point out that simulations used to produce 
predictions for giant arc statistics have often used cosmological parameter values 
that are now disfavored, such as $\sigma_{8} \geq 0.9$. Placing background sources at 
higher redshifts increases the predicted giant arcs counts, but cosmologies with 
lower values of $\sigma_{8}$ should have significantly fewer extremely massive galaxy 
clusters and therefore significantly fewer giant arcs. The ``giant arc problem'' can be 
definitively resolved in the future by combining simulations that use current 
cosmological parameters with the empirically constrained background source redshift 
distribution presented here.

\acknowledgments{MBB acknowledges support from the Illinois Space Grant 
Consortium in the form of a graduate fellowship. JFH acknowledges support provided by 
the Alexander von Humboldt Foundation in the framework of the Sofja Kovalevskaja Award 
endowed by the German Federal Ministry of Education and Research.}

\bibliographystyle{apj}

\clearpage

\end{document}